# Modulation in background music influences sustained attention

Kevin J P Woods, Adam Hewett, Andrea Spencer, Benjamin Morillon, Psyche Loui

**Abstract:** Background music is known to affect performance on cognitive tasks, possibly due to temporal modulations in the acoustic signal, but little is known about how music should be designed to aid performance. Since acoustic modulation has been shown to shape neural activity in known networks, we chose to test the effects of acoustic modulation on sustained attention, which requires activity in these networks and is a common ingredient for success across many tasks. To understand how specific aspects of background music influence sustained attention, we manipulated the rate and depth of amplitude modulations imposed on otherwise identical music. This produced stimuli that were musically and acoustically identical except for a peak in the modulation spectrum that could change intensity or shift location under manipulations of depth or rate respectively. These controlled musical backgrounds were presented to participants (total N = 677) during the sustained attention to response (SART) task. In two experiments, we show performance benefits due to added modulation, with best performance at 16 Hz (beta band) rate and higher modulation depths; neighboring parameter settings did not produce this benefit. Further examination of individual differences within our overall sample showed that those with a high level of self-reported ADHD symptomaticity tended to perform better with more intense beta modulation. These results suggest optimal parameters for adding modulation to background music, which are consistent with theories of oscillatory dynamics that relate auditory stimulation to behavior, yet demonstrate the need for a personalized approach in creating functional music for everyday use.

## Introduction

The ability to sustain attention over time is important for a variety of everyday tasks, but lapses in sustained attention, such as in mind-wandering, can be detrimental to task performance (Smallwood & Schooler, 2006; Esterman et al, 2013; Fortenbaugh et al, 2015; Szpunar et al, 2013). One common strategy to improve performance in everyday tasks is to use background music (Hallam et al, 2002). Music can affect cognition and behavior even when the music is not actively attended (Rauscher, Shaw, & Ky, 1993; Hallam et al, 2002; Husain, Thompson, & Schellenberg, 2002; Schellenberg & Hallam, 2005; Palmiero et al, 2015). Since many tasks don't require hearing, people often use background music, not only to keep themselves entertained, but also to help improve focus, regulate emotion, or improve mood and increase arousal more generally (Thompson, Schellenberg, & Husain, 2001; Schellenberg & Hallam, 2005; Roth & Smith, 2008; Palmiero et al, 2015; Elvers & Steffens, 2017). Given this, surprisingly few studies have manipulated features of music in a controlled manner to uncover relationships between background music and behavior. Instead, work in this area typically compares music to silence (Rauscher, Shaw, & Ky, 1993) or to categorically different stimulation (e.g., noise, tones; Furnham & Strbac, 2002), and is concerned with effects of personality and musical preference (Rentfrow & Gosling, 2003), or demographic predictors (Mullensiefen et al, 2014). These studies ask how music affects people, but the specifics of the music itself are not the issue.

## Control of sound features in background music

How do specific features of music shape behavior? This has been studied largely in terms of categories such as musical genre (Ballard, Dodson, & Bazzini, 1999; Garlin & Own, 2006) and 'musical complexity' (North & Hargreaves, 1995; 1999), which are not easily characterized or quantified. Rather than comparing different 'types' of music, we designed an approach to parametrically manipulate sound streams along a specific acoustic dimension, and compare behavior under conditions where background music is identical but for this manipulation.

Playback volume and tempo alone have been studied in this way, with some studies showing effects of volume and tempo of background music on behavior (Wolfe, 1983; Millman, 1986). However, knowledge of the optimal volume that affects behavior does not readily translate into designing new music for behavioral interventions, since playback volume is not a property of the music per se. Studies comparing tempos have used different pieces of music or used genre as a proxy for tempo (Mayfield & Moss, 1989; Brodsky, 2001). Thus in these studies, tempo was confounded with other acoustic features such as timbre and texture, informational density, and modulation rates, that are in principle independent of tempo.

## Creating controlled background music by manipulating the modulation spectrum

To assess the effect of specific feature dimensions in music on task performance in naturalistic settings, we aimed for a feature of sound that is quantifiable (signal-computable) and which can be set over a range of values with minimal disruption to other aspects of the music. Auditory processing at the cortical level is defined by three main dimensions: the frequency spectrum, the temporal modulation spectrum and the spectral modulation spectrum (Hall et al., 2002, Singh & Theunissen 2003, Schönwiesner & Zatorre 2009). If manipulating the frequency spectrum (roughly, treble-bass balance) one would need to ensure that spectral regions contained redundant information (e.g., by using timbres and pitch ranges that produce only broadband events), or filtering could be confounded with the removal of musical content. Further, there are practical challenges in delivering precise audio spectra over personal listening devices, which can vary widely in frequency response profiles (Møller et al, 1995). This is a difficulty both for experimenters and those who wish to design functional music for widespread use.

Here we thus targeted the amplitude modulation spectrum, by manipulating the rate and depth of intensity modulation. All music contains amplitude modulation (quick changes in loudness) at many rates. Differences in amplitude modulation lead to percepts of 'smooth' or 'rough' sound, and are heard as aspects of timbre, and as sonic texture in music (Ding et al., 2017). The need to quantify acoustic modulation presented a unique challenge in our context. In audio engineering, modulation is quantified in 'percent depth'; however this does not fully describe the result of adding modulation to a complex signal like music. We wanted to measure acoustic modulation at the output of our processing, instead of treating it in terms of the percent depth applied. To do this we turned to a representation used almost exclusively in auditory neuroscience, the modulation spectrum (Fig.1). The modulation spectrum computes amplitude modulation for each auditory channel which is the output of a cochlear filterbank; thus it incorporates knowledge of the auditory periphery and quantifies input to the central nervous system. Although this method of quantifying modulation is a natural solution, to our knowledge the modulation spectrum representation has not previously been used to engineer music. By using audio processing techniques (patent #7674224 awarded to Brain.fm) to concentrate acoustic modulation energy at particular rates, Brain.fm manipulates music in the modulation domain, with the intention of affecting the listener to a much greater and more controllable degree than music normally would.

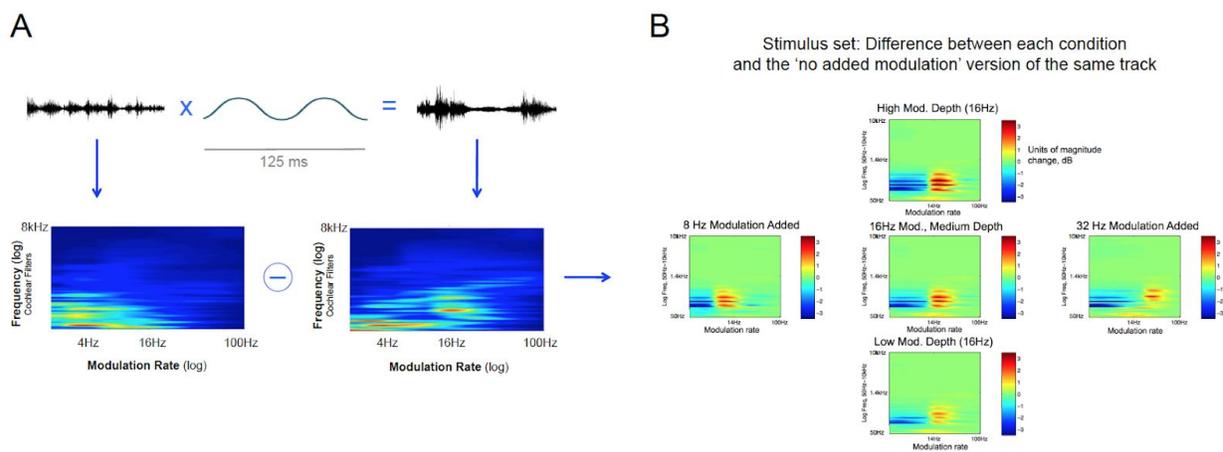

***Figure 1. Controlling modulation in music stimuli.*** *Modulation in sound can be summarized by the modulation spectrum: The vertical axis on these 2D plots depicts audio frequency (0-8kHz). The horizontal axis shows the amplitude modulation spectrum (amplitude fluctuations from 0-100Hz). A) Applying modulation to music results in a modulation spectrum with a peak corresponding to the rate of the added modulation. Upper: The pressure wave is multiplied by a modulator (in this case 16Hz) to produce a modulated signal. Lower: This pressure wave is bandpass filtered, and fluctuation rates in each channel are shown in the modulation spectrum. B) Validation of stimulus manipulations. The stimulus space is illustrated by these panels, each of which shows the difference in modulation spectrum between an unmodulated track and the experimental conditions derived from this track. Rate of added modulation increases moving rightward, while depth increases moving upward.*

To test how acoustic features in background music shape performance, we manipulated the rate and depth of intensity modulations added to otherwise identical background music. We produced background tracks that were musically and acoustically identical except for a peak in the modulation spectrum that would grow or shift under manipulations of depth or rate respectively (Fig.1B). In Figure 1B, the red regions (indicating large difference from the unmodulated condition) thus move rightward (indicating higher modulation rate) from left to right panels, and become darker red (indicating greater modulation depth) from bottom to top. The blue region shows that adding modulation at a particular rate decreases relative modulation energy at other rates, as is necessarily the case when the energy overall is held constant.

The absence of differences elsewhere on the modulation spectrum shows that the experimental conditions were altered in a controlled way, with modulation properties different from the original only as specified (stimulus validation). This level of stimulus control—using the exact same music across conditions apart from a precise manipulation along a well-defined acoustic dimension—is rare in the literature, and we know of no comparable effort to test isolated features of music in a controlled way. Instead, studies of background music generally compare entirely different pieces of music, looking for effects of 'genre' or 'musical complexity' (Furnham and Allass, 1999; Kämpfe et al., 2011; Thompson et al., 2012), terms that are not easily characterized or quantified. Our approach, comparing background music that differs quantifiably in a single acoustic dimension, allows us to attribute any behavioral difference between experimental conditions to this particular aspect of the music.

**Parametrically testing the effects of amplitude modulations**
Having outlined our general approach to studying the effects of amplitude modulation in music, it quickly becomes evident that one can generate infinite conditions of rates and depth modulations, each of which might have a different effect on behavior. To sample this hypothesis space most efficiently, the present study employed a central composite design (a subset of the response surface methodology), rather than a full factorial design (Box & Draper, 2006).

Acoustic amplitude modulations are known to *entrain* neural oscillations, i.e. to induce a selective amplification of neural activity at this frequency. This entrainment occurs along the auditory pathway but also in other cortical networks, such as the attentional network (Besle et al., 2011, Lakatos et al. 2008, Schroeder et al 2009, Calderone et al. 2104). We here hypothesized that it also impacts cognitive processes.

We chose to test rates of 8Hz, 16Hz, and 32Hz for two reasons: First, these fall within ranges of distinct neural oscillatory regimes that are known to have different functions in the brain. Alpha (8-10Hz), Beta (12-25Hz) and Gamma (25-100Hz) are three such ranges of neural oscillatory rhythms, and our experimental conditions using 8Hz, 16Hz, and 32Hz modulation thus fall neatly into each of these oscillatory regimes. Second, these rates were chosen to correspond to note values; we used music at 120bpm so that these modulation rates correspond to 16th, 32nd, and 64th notes. The modulation patterns were also aligned to the metrical grid of the music. This scheme meant that the relationship of the underlying music to the added modulation was as consistent as possible over very different rates. This is desirable in controlling for differences between conditions that arise from intrinsic properties of the underlying (pre-modulation) acoustic signal. For example, a musical event (e.g. a drum hit) transiently amplified by a modulation peak in the 8Hz condition would also be so in the higher-rate conditions. This is only the case because the different modulation rates are aligned to the music and are integer multiples of each other.

We hypothesized that the 16Hz rate would produce effects on performance superior to the no modulation condition, while the other rates would not. This is because beta-band cortical activity is implicated in 'maintenance of the current sensorimotor or cognitive state' (Engel and Fries, 2010), such that increased beta activity is seen when an individual intends to maintain their internal status quo. Beta-band synchrony across the brain is also known to be important for top-down processing in general (Bressler and Richter, 2015) including attentional control (Gross et al., 2004; Lee et al., 2013); for example, a beta-band increase is observed in the hemisphere representing an attended stimulus in spatial attention tasks, resulting in enhanced processing of the attended stimulus (Richter et al., 2018). In contrast, alpha-band enhancement (corresponding to our 8Hz condition) is most often associated with neural inhibition and drowsiness (Klimesch 2012), while gamma-band activity (corresponding to our 32Hz condition) is implicated in bottom-up attention (non-volitional) and seems to operate in transient bursts of synchronized activity in response to sensory input (Jensen et al., 2007). Given this, entraining neural oscillations in the beta band with the 16Hz modulation condition seems most likely to provide a performance benefit.

Depth of modulation refers to how heavily the sound is modulated, rather than the rate of modulation. A maximal depth of modulation would mean that sound energy is reduced to zero at the troughs of the applied modulating waveform, while a very low depth would mean barely-perceptible modulation. While a greater modulation depth is expected to impact neural oscillations more strongly, beyond a point the underlying music suffers aesthetically and the sound becomes distracting due to the sudden changes in loudness over time (auditory salience; Kayser et al., 2005).

**Could rapid modulation be useful for sustained attention?**

The ability to modulate brain rhythms in order to boost task performance has recently been demonstrated (Marshall et al., 2006; Albouy et al., 2017; Andrillon et al., 2017), but only using methods that require specialized hardware (e.g., magnetic stimulation or real-time feedback from electrophysiology or neuroimaging), which renders the methods impractical for widespread use. In contrast, music is accessible and enjoyable, and is also known to entrain brain rhythms (Fujioka et al., 2009; Doelling and Poeppel, 2015). This allows for the possibility that music could also be used to boost performance in a similar way.

But if behavioral effects due to background music involve a wider set of processes, then it is possible for this kind of rapid amplitude modulation to have a positive impact on behavior. For example, the modulation rate we use here (16Hz) is significant for its place in the beta range of neuronal oscillations, which is known to be sensitive to rhythm and meter during music processing (Fujioka et al, 2015). This neuronal entrainment is thought to originate from the motor cortex and be directed towards sensory cortices, thus affecting attention (Morillon & Baillet, 2017). If mechanisms like these underlie the effects of background music, then we might expect to see added modulation in the beta band positively and selectively impacting behavior, specifically behavior that requires sustained attention (Calderone et al, 2014). Furthermore, we might expect that increasing the depth of modulation would result in more benefits to attentive behavior by increasing auditory salience — or the perceptibility of modulation — at least until the modulation depth is so salient that it leads to distraction and annoyance, which outweighs any benefits that the amplitude modulation might have on sustained attention.

To evaluate the participants' level of focus, we use the Sustained Attention to Response Task (SART), which is a gold standard in evaluating lapses of sustained attention, impulsivity, and 'mind-wandering' behavior (Robertson et al, 1997; Helton, 2009; Smilek et al., 2010). In this task, participants respond with a button-press on most trials, but must occasionally withhold their response (effectively, a go/no-go task with infrequency no-go trials). Losing focus (mind-wandering) during this task is indicated by an increase in commission errors, aka false alarm, when participants erroneously push the button on a 'no-go' trial.

In sum, we asked participants to complete the SART in a naturalistic environment, and measured their commission error rates while exposing them to tailor-made background music conditions that systematically and parametrically varied in depth and rate of amplitude modulation, in order to assess the effects of amplitude modulation on sustained attentive behavior.

**For whom might added modulation be useful?**

If effects are indeed mediated by entrainment of neural oscillations, then there may be additional implications. Of particular interest is the possibility that modulations could be most beneficial for individuals with difficulty maintaining sustained attention, such as those with ADHD or related subclinical symptoms of inattention and/or hyperactivity. This seems possible because ADHD is associated with aberrant oscillatory activity, specifically a decreased beta power and increased theta power (Arns et al, 2012), a finding that has motivated neurofeedback as well as rhythmic stimulation strategies for intervention (Calderone et al, 2014). If manipulating the depth of rapid modulations can regulate oscillatory activity through entrainment, then we might expect an interaction between the effects of modulation and self-reported ADHD-like tendencies, because oscillatory activity in the brain is known to differ in this population (Barry et al., 2003; Snyder and Hall, 2006; Ogrim et al., 2012; Arns et al., 2013). To test this we asked our participants to fill out the six-question Adult ADHD Self-Report Scale (ASRS), which is a standard screening tool for ADHD symptoms (Kessler et al., 2005; Adler et al., 2006). This allowed us to compare data between the more and less ADHD-like participants, and to investigate whether t a benefit would be specifically seen among those with tendencies toward attentional deficits.

In addition to ADHD symptoms, we were interested in whether individual differences in the everyday use of music affected participants' sensitivity to modulation rate. Previous studies have shown that people who were introverted were more likely to use music to regulate their own emotions (Chamorro-Premuzic & Furnham, 2007). Based on this, we also incorporated a brief personality inventory (Gosling, Rentfrow, & Swann, 2003), which included a variable of introversion/extroversion in our assessment of the effects of modulation rate on behavior.

## Methods

We carried out two experiments, each with six conditions (Fig. 1B): no-modulation, 8 Hz medium-depth modulation, 16 Hz medium-depth modulation, 32 Hz medium-depth modulation, 16 Hz low-depth modulation, and 16 Hz high-depth modulation. Experiment 1 presented all six conditions to different participant groups, with each condition lasting 15 minutes (i.e. a between-subjects design); each participant heard only one modulation condition. This enabled longer presentations of each condition, thus allowing us to track the effect of each condition over time but limiting the statistical power of comparisons between conditions. In Experiment 2, two groups of participants each heard four conditions, presented for five minutes each in counterbalanced order. The first group of participants heard no-modulation, 8 Hz medium-depth, 16 Hz medium-depth, and 32 Hz medium-depth modulation. The second group heard no-modulation, 16 Hz low-depth, 16 Hz medium-depth, and 16 Hz high-depth modulation. Thus, rate and depth (4 conditions each) were tested on separate groups of participants, but each participant heard all four possible rates or all four possible depths for five minutes each. This limited the duration of each condition, but maximized statistical power in comparing across all the conditions in one dimension of the modulation spectrum, while controlling for intrinsic between-subject differences in performance that are unrelated to our conditions of interest.

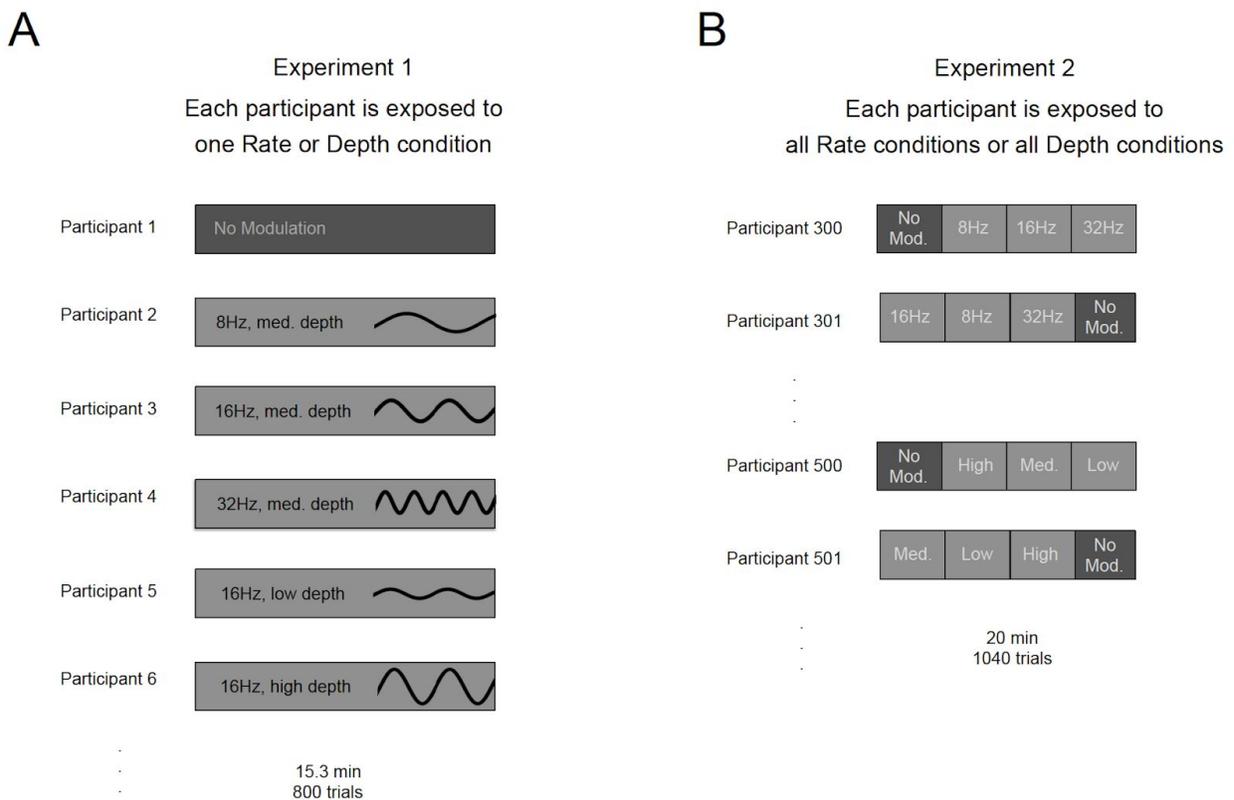

**Figure 2. Design of experiments.** Two experiments tested conditions between and within participants. A) In Experiment 1, each participant heard as background music just one condition, for the entire duration of the experiment (~15 mins). B) In Experiment 2, each participant heard four conditions in blocks of 5 minutes each, with smooth transitions between them. The conditions heard by each participant varied in rate or depth but not both (e.g., no participant heard both the low-depth and fast-rate conditions).

## Participants

We used Amazon's Mechanical Turk platform to recruit and enroll participants. Results were expected to be variable, as they were administered in a naturalistic setting. Thus, our sample size was large (N = 1674 starting sample) to account for relatively large expected sources of variance. Experiment 1 involved 297 participants (166 male, 130 female, 1 other; mean age 36.4). Experiment 2 involved 380 participants (213 male, 168 female; mean age 36.8). In Experiment 1 the conditions (or condition orders) were randomly assigned to participants, such that the number of participants per group was not the same (mean of 49.5, stdv = 6.3; minimum 40, maximum 56).

      To ensure that participants were using headphones as instructed and were exposed to the music throughout the experiment as intended, we administered checks that participants were indeed completing the task while wearing headphones and with the sound on (Woods et al, 2017): At the conclusion of the task, a voice (audio only) gave the participant a number to enter in a text box on the next page. If the audio was off at that time, the response would be blank or incorrect. We were also able to detect headphone use and volume settings, all critical information in auditory experiments. Participants who failed the headphone screening were removed from data analysis. After removing subjects who failed the audio post-check, and accounting for attrition during the study, the final sample was N = 677 usable subjects overall.

## Stimuli and Procedure

The stimuli (background music) were based on two different musical tracks; each had variants created that added amplitude modulation at three rates (8, 16, 32 Hz) and depths (low, medium, high). Modulation depth differences were quantified after processing to account for interactions between the music and modulator. We used the difference between original and processed tracks' modulation spectra (in each cochlear channel; Fig 1) as a metric of applied modulation depth, and set the modulator such that our three depth conditions stepped up evenly in terms of this metric going from low to high depth. In Experiment 2, the transitions between conditions were implemented with smooth crossfades preserving track location (rather than a break and starting the music from the beginning).

      As can be seen in Fig.1B, the applied modulation differences exist predominantly in the low-mid range of the frequency spectrum, with little modulation difference at high frequencies. This was by design, due to aesthetic considerations given the spectrotemporal density of the underlying music: Modulation applied in a broadband manner tended to interact with sparse events in the high frequency regions, which was occasionally annoying (salient). We confined added modulation to lower frequencies by applying it to only the frequency range 200Hz-1kHz.

      Acoustic analysis before and after processing showed that the manipulated stimuli differed from the originals only in the modulation domain, and not in the audio frequency spectrum. Our conditions were therefore identical in terms of musical content and spectral balance ('EQ'), eliminating important confounding factors and ensuring any behavioral differences can be attributed to applied modulation alone.

      Participants provided informed consent as approved by IRB # 120180271 of New England IRB. Users enrolled via Amazon's Mechanical Turk platform. To enroll, users must be over 18 and have normal hearing by self-report. If they chose to participate in our experiment,

they were directed to a cover page with consent documentation and a simple description of the task, followed by a page with payment information. If they still chose to participate, they were told to set their device volume to a low level, and an audio track was played, which they turned up to a comfortable volume. In this way, the volume was determined by the listener, and was limited by the consumer hardware and software they were using (headphones, soundcard). At this point they were directed to a screen with instructions, and could begin the task when ready.

Participants completed a sustained attention to response task (SART) (Robertson et al, 1997; Helton, 2009; Smilek et al., 2010). In this task, a single digit ranging from 0 to 9 appeared on the screen for each trial. Each digit was presented for 250 ms followed by a 900 ms mask, resulting in a 1150 ms inter-trial interval. Participants' task was to respond to any digit except for 0. Instructions for the task were as follows: "Numbers will appear on the screen. If you see 1-9, hit any key; if you see 0 do not hit any key." Participants were allowed to quit at any time, and were told this at the outset. Participants were paid at a rate of $0.01 per correct response and -$0.10 per commission error (misses were given no pay; $0.00), resulting in an average of ~$12/hr for the overall task. Participants are not penalized in any way for leaving.

Participants were told that the background music was unrelated to the task, but that they should listen to the background music in order to control the acoustic environment across participants. Participants were not told anything about the modulation, or the fact that there were different types of background music. In Experiment 1, participants were told the experiment would run essentially forever and they should leave as they wished. In reality the music stopped only after 4000 trials (78 minutes). When we did this, attrition reached 50% after 20 minutes, considerably inflating the error in measurement. To limit our analysis to trials with a sufficient sample size, we chose to analyze just the first 20% of the trials (~15 minutes, 800 trials) in Experiment 1. In Experiment 2, participants were told the experiment would run for only 20 minutes, and they should try to complete the entire 20 minutes of testing. In this experiment the music smoothly transitioned between the different modulation conditions while the task continued without any interruption; participants were not told anything about the changes in modulation, and the music stopped when the task was over (i.e. the 1044 trials were complete).

**Data Analysis**

Raw data from Amazon's Mechanical Turk were exported to Matlab for analysis. The dependent variable was the rate of commission errors, and the independent variables were modulation rate and modulation depth. To test the effects of individual differences in attention difficulties, as quantified by the ASRS (Kessler et al, 2005), we did a median split on all subjects' ASRS scores, yielding a high-ASRS group (i.e. those with more ADHD-like symptoms) and a low-ASRS group (those with less ADHD-like symptoms). To test the effects of individual differences in extroversion as quantified by the Ten-Item Personality Inventory (Gosling et al, 2003), we did a median split on all subjects' extroversion scores to yield an extrovert group and an introvert group. These two factors, ASRS (high vs. low) and extroversion (extrovert vs. introvert), were additionally incorporated as between-subjects factors in our overall statistical model.

In experiment 1, where participants were continuously exposed to each modulation condition for over 15 minutes, an additional variable of interest was time to onset of the effect of modulation. This was because we had no *a priori* reason to believe that any effect of modulation on commission error rates would begin as soon as the music started; rather we expected that the modulation manipulation would require some time to show an effect relative to the no-modulation control condition. To compare the commission error rate at each time point between different modulation conditions, we averaged the commission error rate over time bins, with 200 trials per bin. The 200-trial time bin (~4 minutes) was chosen because it resulted in the most stable commission error rate across bins. The average commission error rate for each time bin was then compared between modulation conditions and the no-modulation condition. Importantly, to see whether the effect of modulation was different between groups of subjects, a repeated-measures ANOVA was done on the dependent variable of commission error rate, with the repeated-measures variable of time window, and the between-subjects factors of rate (4 levels: no-modulation, 8 Hz, 16 Hz, 32 Hz) or depth (4 levels: no-modulation, low depth, medium depth, high depth), and the between-subjects factor of ASRS (high vs. low), and Extraversion (high vs. low).

For experiment 2, a mixed-effects ANOVA was run on the dependent variable of commission error rate, with the within-subjects factors of rate (4 levels: no-modulation, 8 Hz, 16 Hz, 32 Hz), and the between-subjects factors of ASRS (high vs. low) and extroversion (high vs. low). A second mixed-effects ANOVA was run on the dependent variable of commission error rate, with the within-subjects factors of depth (4 levels: no-modulation, low depth, medium depth, high depth), and the between-subjects factors of ASRS (high vs. low) and extroversion (high vs. low).

## Results

### Modulation Rate

Figure 5 shows commission error rates for each modulation condition compared against no-modulation control, as a function of time (Experiment 1: Figure 5A) and across different conditions (Experiment 2: Figure 5B). Performance under background music with no added modulation (the baseline control condition) is plotted in black. Performance under the same music when modulation is added at various rates is represented by the dashed grey lines (Experiment 1: Figure 5A) or grey bars (Experiment 2: Figure 5B), with our 16Hz condition shown in magenta.

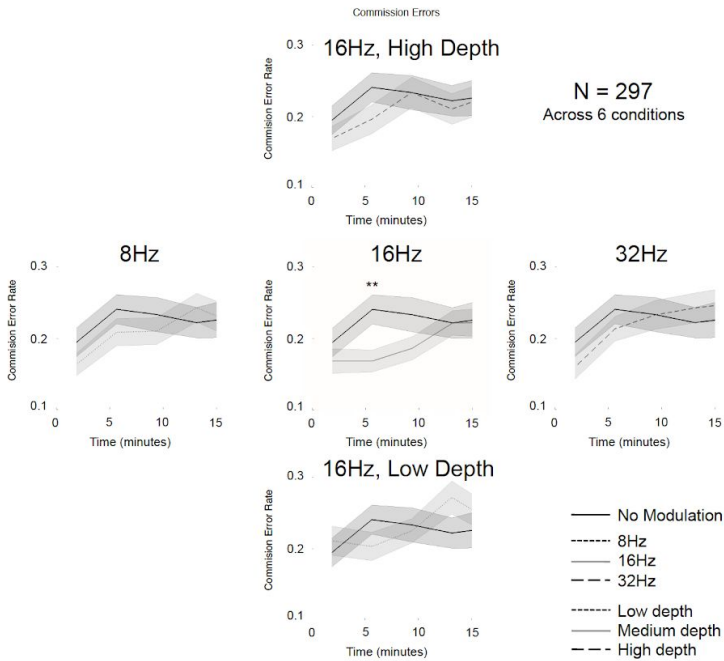

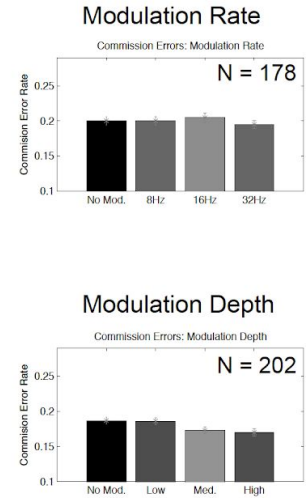

**Figure 3. Commission error rates under background music identical but for the rate or depth of added modulation.** A) Experiment 1 (conditions between subjects). Conditions organized as in Fig1B (Rate increasing upward, depth rightward). The unique condition in each panel is shown in light gray (dashed). The 'No Modulation' control condition (solid black) is shown in each case, for comparison. Performance is taken over windows of 200 trials. N=297 participants randomly distributed across conditions. B) Experiment 2 (conditions within subjects). One set of subjects heard the rate conditions (N=178); a different group of subjects heard the depth conditions (N=202). Conditions appeared in 5-minute blocks, in random order. In all cases, error bars depict +/-1 standard error of the mean. **p < .005, surviving corrections for comparisons across 4 bins.

### Experiment 1: Between-subjects effects of modulation rate

Since there were four sequential time windows in each condition, with each time window corresponding to the averaged commission error rate over 200 trials, we used a repeated-measures design where time window was a within-subjects variable, and modulation rate (no modulation, 8 Hz, 16 Hz, and 32 Hz modulations) and ASRS (high vs. low) were between-subject variables. The effect of time window was highly significant ($F_{(3,191)} = 13.918$, $p < .001$), and a followup linear contrast across the four time windows was highly significant ($F_{(1,193)} = 39.506$, $p < .001$), indicating that commission error rates increased over time; this increase in commission errors over time is a well-replicated sign of failures in sustained attention (Fortenbaugh et al, 2015).

Importantly, there was a significant interaction between time window and modulation rate: $F_{(3,193)}$ = 3.349, p = .02), indicating that the rate of amplitude modulation affected performance differently over time. Followup contrast tests showed a significant quadratic contrast in this time window by modulation rate interaction ($F_{(3,193)}$ = 2.696, p = .047). This indicates that the modulation rate of the background music affected the u-shaped or inverse u-shaped performance curve over time: specifically, participants showed better performance in the middle time windows (200-600 trials after the start of the task) when listening to the middle (16 Hz) modulation rate condition, compared to other modulation rate conditions.

The difference in commission errors between the unmodulated and 16Hz modulation conditions (black vs magenta) was statistically significant in the second time-bin ($t_{(89)}$ = 3.01, p < 0.005), surviving Bonferroni correction for possible comparisons across four time-bins. A difference also in the third time-bin is visible but not statistically significant ($t_{(89)}$ = 1.66, p = 0.10). The other modulation rates (8Hz and 32Hz) produced commission errors that were not significantly different from the no-modulation control condition, but were also not significantly different from the 16Hz condition.

In general, commission errors in the SART (regardless of modulation condition) increased by about one-third (from 18% to 24%) over the course of the experiment, reflecting the beginning of mind-wandering or the onset of failures in sustained attention that generally occurred within the first 5 minutes, as the task became boring to the participant. Importantly, adding moderate modulation to the music at a 16Hz rate (magenta in Fig 1A) delayed this decline until later in the experiment, as shown by a slower increase in commission error rate over time. This pattern of performance in commission errors suggests a behavioral advantage for the 16Hz rate over the no-modulation control and the other modulation rates.

### Experiment 2: Within-subjects effects of modulation rate

For Experiment 2, a mixed-effects ANOVA on the dependent variable of commission error rate, with the within-subject independent variable of modulation rate (4 levels: no-modulation, 8 Hz, 16 Hz, 32 Hz), and the between-subjects factors of ASRS (high vs. low) and extroversion (high vs. low), showed a significant between-subjects effect of ASRS ($F_{(1,173)}$ = 6.703, p = 0.01), where high-ASRS participants had more commission errors than low-ASRS participants. There was also a significant interaction between the factors of ASRS and Extroversion ($F_{(1,173)}$ = 5.491, p = 0.02): the high-ASRS introverted group had lower commission error rate for the 16 Hz modulation condition than for the no-modulation condition, whereas the other groups showed higher or no difference in commission error rates across conditions. In other words, the predicted effect of lowest commission error rate for the 16 Hz modulation condition was observed only among introverts with high ADHD symptoms. Importantly, followup within-subject contrasts tests showed a significant interaction between ASRS and the quadratic contrast of modulation rate (Rate x ASRS: $F_{(1,173)}$ = 4.591, p = 0.034; Fig. 4B). This indicates that for the high-ASRS group, the middle frequency modulation condition (16 Hz) resulted in the best performance (i.e. lowest commission error rates), whereas for the low-ASRS group, the same condition resulted in the worst performance (highest commission errors). No other effects were significant at the p < .05 level.

In follow-up analyses, we were additionally interested in whether the order of different stimulation conditions affected the results, since in Experiment 1 we observed that the significant effect of modulation rate on performance occurred after 200 trials. To follow up on the effect of changes in modulation rate over time, the order of each condition was entered as a variable in a linear mixed effects model. There was a significant effect of condition order ($F(3,177) = 26.8$, $p < .001$), with the first two conditions having a significant effect (1st condition: $t(177) = -8.27$, $p < .001$; 2nd condition: $t(177) = -5.108$, $p < .001$). Those who started on the no-modulation condition performed better in the first condition than those who started on other modulation conditions, suggesting that the amplitude modulation works best if it is introduced gradually.

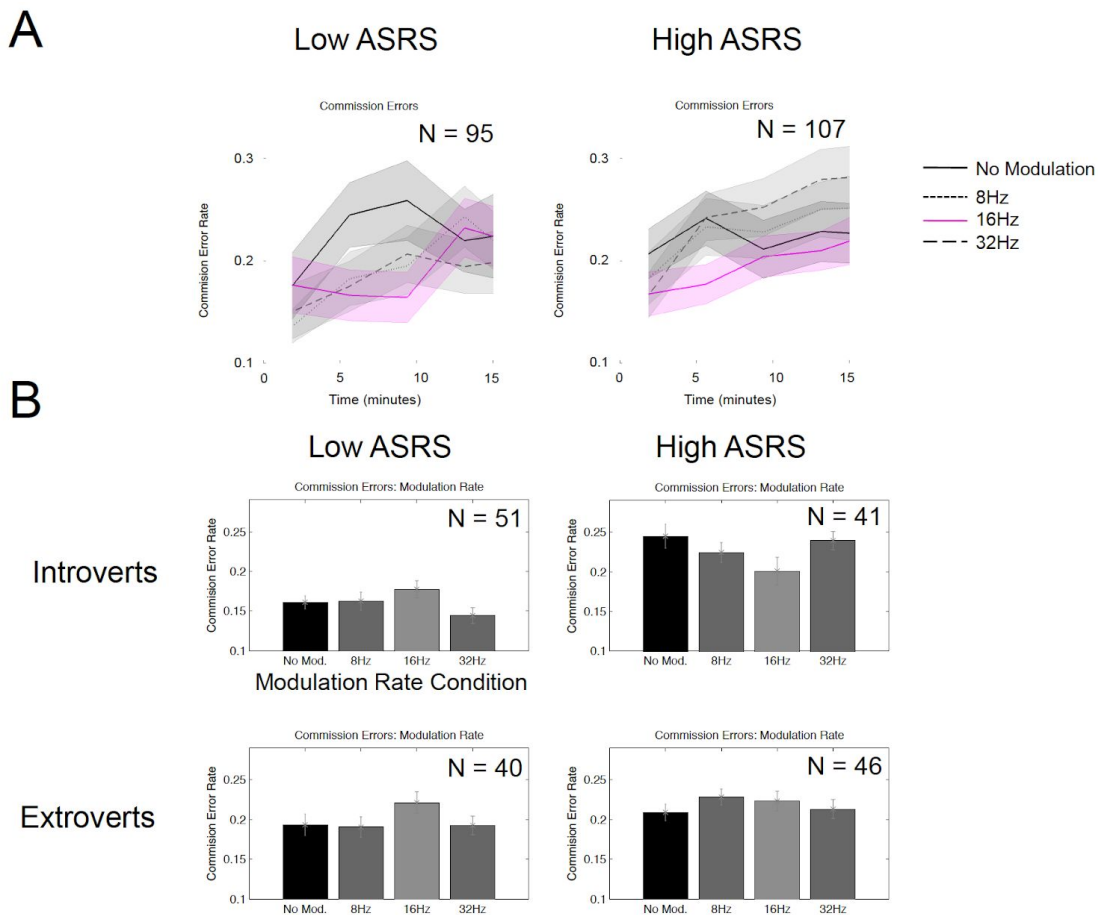

**Figure 4. Modulation Rate results analysed by participant subgroups.** A) Experiment 1 (between-subjects design). 16Hz is shown in magenta instead of light gray. No Modulation and 16Hz are the same data as appear in Fig.3A. Gray lines depict the 8Hz and 32Hz conditions (fine and coarse dashes respectively). B) Experiment 2 (within-subjects design), showing participants split by ASRS and Extroversion.

**Modulation Depth**

Greater modulation depth is expected to impact neural oscillations more strongly, and so we hypothesized that an intermediate depth would be most useful. Alternatively, it could turn out that musicality is irrelevant compared to entrainment, and a very high depth is most effective. Or, it could turn out that our starting modulation depth was already too salient, in which case we would expect that a lower depth produces better performance.

### Experiment 1: Between-subjects effects of modulation depth

For the modulation depth manipulations, same as in the rate experiment, there was a highly significant effect of time window ($F(3,176) = 6.477$, $p < .001$), with a highly significant linear contrast ($F(1,178) = 19.083$, $p < .001$), indicating that participants generally performed worse over time. Importantly, there was a significant time window by modulation depth interaction ($F(3,178) = 6.642$, $p < .001$), and follow-up contrast tests showed a significant quadratic contrast in the time window by depth interaction ($F(3,178) = 6.384$, $p < .001$). This indicates that some select modulation depth conditions affected participants' performance for the middle time windows relative to the first and last time windows. Specifically, participants performed better on the modulation conditions than on the no-modulation condition, with performance being best on the medium modulation depth condition. This suggests that any behavioral advantage from adding modulation to music requires a moderate depth — not too high and not too low.

Between-subjects contrasts also showed a significant effect of ASRS ($F(1,178) = 6.264$, $p = 0.013$), with high-ASRS participants performing worse than low-ASRS participants. There was also a marginally significant interaction between modulation depth and ASRS ($F(3,178) = 2.596$, $p = 0.05$), indicating that the effect of modulation depth was slightly stronger among participants who scored highly on the ASRS, i.e. those reporting more ADHD symptoms (Fig. 5A). No other main effects or interactions were significant.

### Experiment 2: Within-subjects effects of modulation depth

A mixed effects ANOVA on the dependent variable of commission error rate, with the within-subjects factor of modulation depth (no modulation, low, medium, and high modulation depth) and between-subjects factors of ASRS (high vs. low) and extraversion (extrovert vs. introvert), showed a significant effect of modulation depth ($F(3,196) = 2.98$, $p = .032$, Fig2B bottom). A follow-up contrast test showed a significant linear contrast ($F(1,198) = 7.518$, $p = .007$), indicating that the heavier modulation depths resulted in better performance. There was also a significant between-subjects effect of ASRS ($F(1,198) = 6.811$, $p = 0.01$), with the high-ASRS group showing more commission errors (Fig. 5B). No other contrasts were statistically significant at the $p < 0.05$ level.

We additionally tested for the effects of the order of different stimulation depths. The effect of condition order was again significant ($F(3,202) = 8.00$, $p < .001$), with the first two conditions having a significant effect (1st condition: $t(202) = 4.4$, $p < .01$; 2nd condition: $t(202) = -2.85$, $p < .01$). Those who started on the no-modulation condition performed better in the first condition than those who started on other modulation conditions, suggesting that performance is best when the amplitude modulation is gradually introduced.

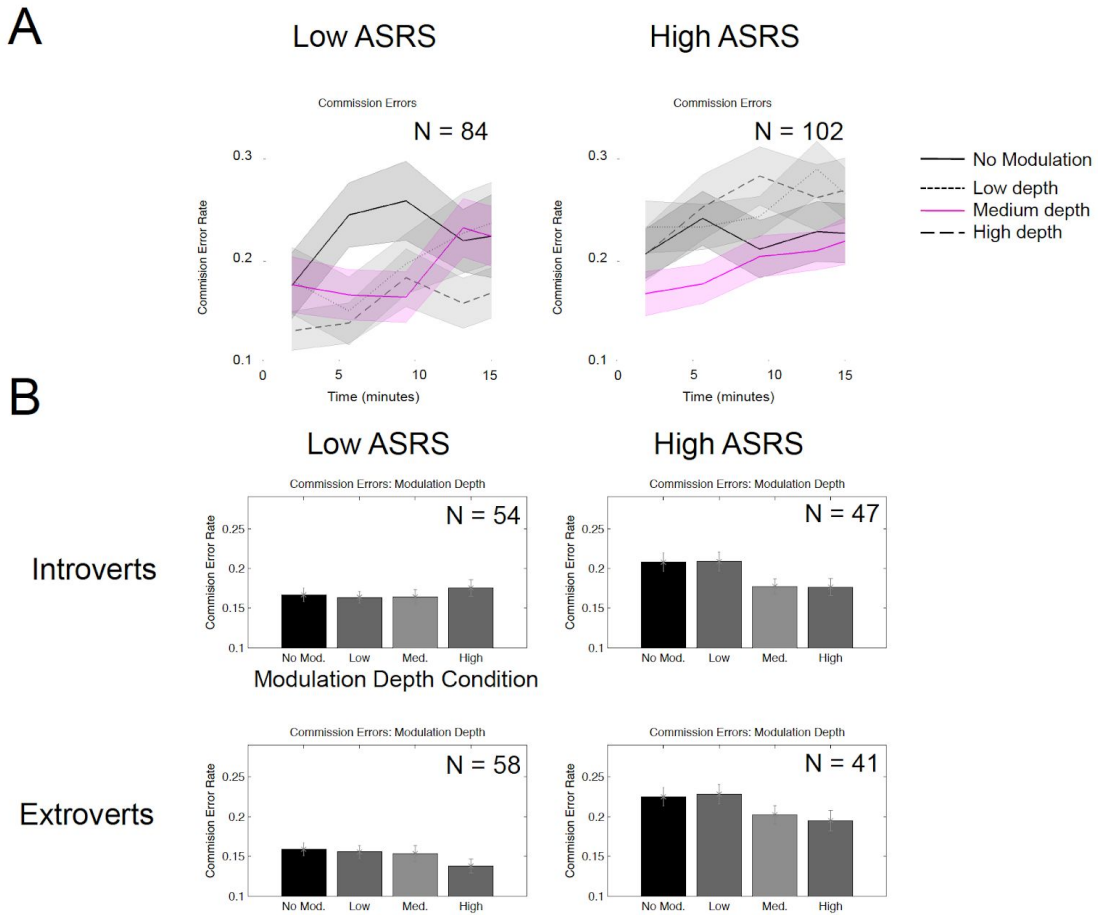

**Figure 5. Modulation Depth results analysed by participant subgroups.** A) Experiment 1 (between-subjects design). Medium depth is shown in magenta instead of light gray. Black (No Modulation) and Magenta (medium depth) here are the same data as appear in Fig.3A. Light gray lines depict the Low- and High-depth conditions (fine and coarse dashes respectively). B) Experiment 2 (within-subjects design), showing participants split by ASRS and Extroversion.

## Discussion

Background music has long been used to aid sustained performance on cognitive tasks, and while previous studies have identified beneficial effects of music on mood and arousal, few research programs have explicitly manipulated well-defined acoustic properties of music in order to test its effect on behavior. Here, using the most rigorously defined and controlled acoustic manipulations to date, we find that the rate and depth of amplitude modulation in music can benefit cognitive task performance, as indexed by decreased commission error rates in the sustained attention to response task (SART). These results have implications for music therapy and for sound design, by highlighting the use of amplitude modulations as an active ingredient for receptive music interventions for those with neurological and/or psychiatric disorders, as well as for those with no diagnosed deficit that might nonetheless experience subclinical symptoms of inattention and hyperactivity.

Given extended duration of stimulation, the most beneficial condition was the 16 Hzamplitude modulation at a moderate depth; this affected behavior in both between-subjects (exp. 1) and within-subjects (exp. 2) experiments, with the largest effect observed in the between-subjects experiment approximately 5 minutes after the onset of the music. The between-subjects experiment showed a significantly lower commission error rate at the 5 minute time-point in the medium depth 16 Hz amplitude modulation condition compared to the no-modulation condition. None of the other amplitude modulation conditions (8 Hz, and 32 Hz, 16 Hz low depth and 16 Hz high depth) showed the same effect (Fig. 3A). Furthermore, the significant difference was only observed at the second and third 200-trial time window, i.e. from 4 to 12 minutes after the onset of the modulation. This suggests that the effect of beta band modulation on commission error rates is temporally specific: the beneficial effect unfolds several minutes after the onset of the amplitude modulation, and lasts for approximately 10 minutes.

The return to baseline of the effect after the third time window may be attributable to fatigue or habituation, or it may reflect the relative ease and simplicity of the task. While future studies are needed to tease apart these possible explanations for the temporal sensitivity of the effect, results from the within-subjects experiment may offer some clues. In the within-subjects experiment, participants were exposed to a new amplitude modulation condition (in counterbalanced order) every 5 minutes for four different conditions, without any apparent break or silence in between. Participants were not told that there would be changes in the amplitude modulation patterns throughout the overall 20-minute experiment. This much more subtle manipulation abolished the effect of amplitude modulation on commission errors. Since the duration of each amplitude modulation condition was much shorter in the within-subjects experiment than in the between-subjects experiment, the fact that we observed significant individual differences in sensitivity to amplitude modulation only in the between-subjects experiment provides further support for the idea that the effects of amplitude modulation take several minutes to unfold, and the individual differences in sensitivity to modulation depth can be amplified over time. Furthermore, we observe individual differences in the effects of amplitude modulation on commission error rates in the SART: the effect of the beta-band modulation was strongest and most beneficial for those with moderate-to-high ADHD symptoms, and for individuals who self-identify as introverts.

High modulation depth elicited the best performance in the between-subjects experiment, but not in the within-subject experiment. While further work is needed to reconcile the between-subjects and within-subjects results, one clue comes from the fact that participants heard each modulation condition for less time in the within-subjects experiment than in the between-subjects experiment. It may be the case that the effects of heavy modulation depth could only last for a short duration before fatigue or annoyance from the highly salient modulation sets in and affects behavior. In other words, the adverse effects of extremely salient modulations may be mitigated by presenting the heavily applied amplitude modulation in short bursts.

Participants reporting more ADHD-like qualities were particularly sensitive to the rate and depth of modulation; while the parameters we hypothesized would work, did work, high ASRS-participants' performance suffered much more so than the low-ASRS participants if modulation was too fast: 32Hz amplitude modulation led to worse performance than no-modulation in the high-ASRS participants, in both the within-subjects and between-subjects experiments.

While the fastest modulation rate was clearly detrimental for high-ASRS participants, the effects of modulation depth was less clear. Results from the modulation depth manipulations for high-ASRS participants differed between the within-subjects and between-subjects experiments. In the within-subjects experiment, high modulation depth led to the best performance, resulting in the lowest commission error rate of all conditions for this group. In contrast, in the between-subjects experiment, high modulation depth led to the worst performance, with the highest commission error rate. One possible explanation for this discrepancy is that the high-ASRS participants habituate or fatigue more quickly in response to a single rate or depth of modulation; instead they may require relatively quick changes in modulation in order to function more optimally, and closer to the low-ASRS participants. In that regard, it is noteworthy that individual differences in commission error rates overall were highly robust in showing that high-ASRS participants committed more commission errors than low-ASRS participants. Even though the ASRS is a simple self-report survey with only six items (Kessler et al., 2005), it clearly distinguishes between more symptomatic and less symptomatic subgroups of participants that respond differently to the same cognitive task, and require different types of acoustic stimulation to aid their function.

Music with these added modulations is likely to synchronize neural activity, at rates higher than would normally appear in music. This rapid phase-locked activity could be widespread across the brain rather than confined to sensory areas, supporting a mechanism for the observed behavioral effects: That rapid acoustic modulation can shape the global (inter-areal) oscillatory regimes controlling top-down behavior. Neuroimaging studies are starting to show how the functional networks involved can be affected by music (Doelling et al., 2019), but further work is needed to show how global synchronized activity might affect the brain to aid focus.

**Limitations**

The core question we have yet to answer is why shaping acoustic input in this way can improve performance. What exactly are the mechanisms linking neuronal oscillations to behavior? Our results thus far support a picture where widespread neural synchrony induced by music can ultimately aid focus, but this could be happening in multiple ways. Beta was targeted for its role in task-maintenance and global (inter-areal) communication (Engel and Fries, 2010; Bressler and Richter, 2015; Richter et al., 2018), but beta-rate synchrony is also found in other roles, in perception (Fujioka et al., 2009, 2015) and motor output (Tzagarakis et al., 2010; Davis et al., 2012). Our work thus far cannot rule out the possibility that the behavioral effects are due to these other systems (sensory-motor, rather than cognitive). This matters because targeting a particular system may work for one person and not another. Understanding the mechanism will help explain individual variability, predict efficacy, and allow us to set parameters for individual needs. Ongoing neuroimaging work will help by giving a fuller picture of the functional networks involved, as will behavioral experiments that load differently on perception, cognition, and action to tease apart the origins of the effects reported here.

## References

Adler, L.A., Spencer, T., Faraone, S.V., Kessler, R.C., Howes, M.J., Biederman, J., and Secnik, K. (2006). Validity of Pilot Adult ADHD Self- Report Scale (ASRS) to Rate Adult ADHD Symptoms. Ann. Clin. Psychiatry *18*, 145–148.

Albouy, P., Weiss, A., Baillet, S., and Zatorre, R.J. (2017). Selective Entrainment of Theta Oscillations in the Dorsal Stream Causally Enhances Auditory Working Memory Performance. Neuron *94*, 193-206.e5.

Andrillon, T., Pressnitzer, D., Léger, D., and Kouider, S. (2017). Formation and suppression of acoustic memories during human sleep. Nat. Commun. *8*, 179.

Angel, L.A., Polzella, D.J., and Elvers, G.C. (2010). Background Music and Cognitive Performance. Percept. Mot. Skills *110*, 1059–1064.

Arns, M., Conners, C.K., and Kraemer, H.C. (2013). A Decade of EEG Theta/Beta Ratio Research in ADHD: A Meta-Analysis. J. Atten. Disord. *17*, 374–383.

Atlas, L., & Janssen, C. (2005, March). Coherent modulation spectral filtering for single-channel music source separation. In *Proceedings.(ICASSP'05). IEEE International Conference on Acoustics, Speech, and Signal Processing, 2005.* (Vol. 4, pp. iv-461). IEEE.

Atlas, L., Li, Q., & Thompson, J. (2004, May). Homomorphic modulation spectra. In 2004 IEEE International Conference on Acoustics, Speech, and Signal Processing (Vol. 2, pp. ii-761). IEEE.

Barry, R.J., Clarke, A.R., and Johnstone, S.J. (2003). A review of electrophysiology in attention-deficit/hyperactivity disorder: I. Qualitative and quantitative electroencephalography. Clin. Neurophysiol. *114*, 171–183.

Besle, J., Schevon, C. A., Mehta, A. D., Lakatos, P., Goodman, R. R., McKhann, G. M. & Schroeder, C. E. (2011). Tuning of the human neocortex to the temporal dynamics of attended events. *Journal of Neuroscience*, *31*(9), 3176-3185.

Bressler, S.L., and Richter, C.G. (2015). Interareal oscillatory synchronization in top-down neocortical processing. Curr. Opin. Neurobiol. *31*, 62–66.

Buzsáki, G., and Draguhn, A. (2004). Neuronal Oscillations in Cortical Networks. Science *304*, 1926–1929.

Calderone, D.J., Lakatos, P., Butler, P.D., and Castellanos, F.X. (2014). Entrainment of neural oscillations as a modifiable substrate of attention. Trends Cogn. Sci. *18*, 300–309.

Cole, S.R., and Voytek, B. (2017). Brain Oscillations and the Importance of Waveform Shape. Trends Cogn. Sci. *21*, 137–149.

Danckert, J., and Merrifield, C. (2018). Boredom, sustained attention and the default mode network. Exp. Brain Res. *236*, 2507–2518.

Davis, N.J., Tomlinson, S.P., and Morgan, H.M. (2012). The Role of Beta-Frequency Neural Oscillations in Motor Control. J. Neurosci. *32*, 403–404.

Ding, N., Patel, A.D., Chen, L., Butler, H., Luo, C., and Poeppel, D. (2017). Temporal modulations in speech and music. Neurosci. Biobehav. Rev. *81*, 181–187.

Doelling, K.B., and Poeppel, D. (2015). Cortical entrainment to music and its modulation by expertise. Proc. Natl. Acad. Sci. *112*, E6233–E6242.

Elvers, P., & Steffens, J. (2017). The Sound of Success: Investigating Cognitive and Behavioral Effects of Motivational Music in Sports. *Front Psychol, 8*, 2026.

Engel, A.K., and Fries, P. (2010). Beta-band oscillations—signalling the status quo? Curr. Opin. Neurobiol. *20*, 156–165.

Fujioka, T., Trainor, L.J., Large, E.W., and Ross, B. (2009). Beta and Gamma Rhythms in Human Auditory Cortex during Musical Beat Processing. Ann. N. Y. Acad. Sci. *1169*, 89–92.

Fujioka, T., Ross, B., and Trainor, L.J. (2015). Beta-Band Oscillations Represent Auditory Beat and Its Metrical Hierarchy in Perception and Imagery. J. Neurosci. *35*, 15187–15198.

Furnham, A., and Allass, K. (1999). The influence of musical distraction of varying complexity on the cognitive performance of extroverts and introverts. Eur. J. Personal. *13*, 27–38.

Furnham, A., Trew, S., and Sneade, I. (1999). The distracting effects of vocal and instrumental music on the cognitive test performance of introverts and extraverts. Personal. Individ. Differ. *27*, 381–392.

Gross, J., Schmitz, F., Schnitzler, I., Kessler, K., Shapiro, K., Hommel, B., and Schnitzler, A. (2004). Modulation of long-range neural synchrony reflects temporal limitations of visual attention in humans. Proc. Natl. Acad. Sci. *101*, 13050–13055.

Hall, D. A., Johnsrude, I. S., Haggard, M. P., Palmer, A. R., Akeroyd, M. A., & Summerfield, A. Q. (2002). Spectral and temporal processing in human auditory cortex. *Cerebral Cortex*, *12*(2), 140-149.

Helton, W.S. (2009). Impulsive responding and the sustained attention to response task. J. Clin. Exp. Neuropsychol. *31*, 39–47.

Jensen, O., Kaiser, J., and Lachaux, J.-P. (2007). Human gamma-frequency oscillations associated with attention and memory. Trends Neurosci. *30*, 317–324.

Kämpfe, J., Sedlmeier, P., and Renkewitz, F. (2011). The impact of background music on adult listeners: A meta-analysis. Psychol. Music *39*, 424–448.

Kayser, C., Petkov, C.I., Lippert, M., and Logothetis, N.K. (2005). Mechanisms for Allocating Auditory Attention: An Auditory Saliency Map. Curr. Biol. *15*, 1943–1947.

Kessler, R.C., Adler, L., Ames, M., Demler, O., Faraone, S., Hiripi, E., Howes, M.J., Jin, R., Secnik, K., Spencer, T., et al. (2005). The World Health Organization adult ADHD self-report scale (ASRS): a short screening scale for use in the general population. Psychol. Med. *35*, 245–256.

Klimesch, W. (2012). Alpha-band oscillations, attention, and controlled access to stored information. *Trends in cognitive sciences*, *16*(12), 606-617.

Lakatos, P., Karmos, G., Mehta, A. D., Ulbert, I., & Schroeder, C. E. (2008). Entrainment of neuronal oscillations as a mechanism of attentional selection. *science*, *320*(5872), 110-113.

Lee, J.H., Whittington, M.A., and Kopell, N.J. (2013). Top-Down Beta Rhythms Support Selective Attention via Interlaminar Interaction: A Model. PLOS Comput. Biol. *9*, e1003164.

Marshall, L., Helgadóttir, H., Mölle, M., and Born, J. (2006). Boosting slow oscillations during sleep potentiates memory. Nature *444*, 610–613.

Møller, H., Hammershøi, D., Jensen, C. B., & Sørensen, M. F. (1995). Transfer characteristics of headphones measured on human ears. *Journal of the Audio Engineering Society, 43*(4), 203-217.

Ogrim, G., Kropotov, J., and Hestad, K. (2012). The quantitative EEG theta/beta ratio in attention deficit/hyperactivity disorder and normal controls: Sensitivity, specificity, and behavioral correlates. Psychiatry Res. *198*, 482–488.

Palmiero, M., Nori, R., Rogolino, C., D'Amico, S., & Piccardi, L. (2015). Situated navigational working memory: the role of positive mood. *Cogn Process, 16 Suppl 1*, 327-330.

Richter, C.G., Coppola, R., and Bressler, S.L. (2018). Top-down beta oscillatory signaling conveys behavioral context in early visual cortex. Sci. Rep. *8*, 6991.

Roth, E. A., & Smith, K. H. (2008). The Mozart effect: evidence for the arousal hypothesis. *Percept Mot Skills, 107*(2), 396-402.

Schroeder, C. E., & Lakatos, P. (2009). Low-frequency neuronal oscillations as instruments of sensory selection. *Trends in neurosciences, 32*(1), 9-18.

Schönwiesner, M., & Zatorre, R. J. (2009). Spectro-temporal modulation transfer function of single voxels in the human auditory cortex measured with high-resolution fMRI. *Proceedings of the National Academy of Sciences, 106*(34), 14611-14616.

Shirer, W.R., Ryali, S., Rykhlevskaia, E., Menon, V., and Greicius, M.D. (2012). Decoding Subject-Driven Cognitive States with Whole-Brain Connectivity Patterns. Cereb. Cortex *22*, 158–165.

Singh, N. C., & Theunissen, F. E. (2003). Modulation spectra of natural sounds and ethological theories of auditory processing. The Journal of the Acoustical Society of America, 114(6), 3394-3411.

Smilek, D., Carriere, J.S.A., and Cheyne, J.A. (2010). Failures of sustained attention in life, lab, and brain: Ecological validity of the SART. Neuropsychologia *48*, 2564–2570.

Snyder, S.M., and Hall, J.R. (2006). A Meta-analysis of Quantitative EEG Power Associated With Attention-Deficit Hyperactivity Disorder. J. Clin. Neurophysiol. *23*, 441.

Thompson, W.F., Schellenberg, E.G., and Letnic, A.K. (2012). Fast and loud background music disrupts reading comprehension. Psychol. Music *40*, 700–708.

Tzagarakis, C., Ince, N.F., Leuthold, A.C., and Pellizzer, G. (2010). Beta-Band Activity during Motor Planning Reflects Response Uncertainty. J. Neurosci. *30*, 11270–11277.

Ward, L.M. (2003). Synchronous neural oscillations and cognitive processes. Trends Cogn. Sci. *7*, 553–559.

Woods, K.J.P., Siegel, M.H., Traer, J., and McDermott, J.H. (2017). Headphone screening to facilitate web-based auditory experiments. Atten. Percept. Psychophys. *79*(7), 2064-2072.